%% file: main.tex
\documentclass[journal]{IEEEtran}
\input{packageSetup}

\usepackage{wasysym}
\usepackage{diagbox}
\usepackage{tabularx} 
\usepackage{subfigure}
\usepackage{caption}
\usepackage{color}
\usepackage{makecell}
\usepackage{booktabs}
\usepackage{multirow}
\usepackage{eso-pic}

\begin{document}

\title{Zero-Knowledge Federated Learning: A New Trustworthy and Privacy-Preserving Distributed Learning Paradigm}

\author{Taotao~Wang,~\emph{Member,~IEEE},~Yuxin~Jin,~Qing~Yang,~\emph{Member,~IEEE},~Yihan~Xia,~Long~Shi,~\emph{Senior~Member,~IEEE},~and~Shengli~Zhang,~\emph{Senior~Member,~IEEE}

\thanks{T.~Wang,,~Y.~Jin,~Q.~Yang,~Y.~Xia~and~S.~Zhang are with the College of Electronics and Information Engineering, Shenzhen University, Shenzhen, Guangdong Province, PRC, e-mail: {ttwang@szu.edu.cn}, {jinyuxin2022@email.szu.edu.cn}, {yang.qing@szu.edu.cn}, {xiayihan2023@email.szu.edu.cn}, {zsl@szu.edu.cn}. L. Shi is with the School of Electronic and Optical Engineering, Nanjing University of Science and Technology, Nanjing 210094, China, e-mail: longshi@njust.edu.cn.}

}

\maketitle

\maketitle

\AddToShipoutPictureBG*{
	\AtPageLowerLeft{%
		\setlength\unitlength{1in}%
		\put(0,0.25){%
			\makebox(\paperwidth,0)[c]{
				\parbox{0.9\paperwidth}{\centering\footnotesize 
					\copyright~2026 IEEE. Personal use of this material is permitted. Permission from IEEE must be obtained for all other uses, in any current or future media, including reprinting/republishing this material for advertising or promotional purposes, creating new collective works, for resale or redistribution to servers or lists, or reuse of any copyrighted component of this work in other works.
				}%
			}%
		}%
	}%
}

\begin{abstract}

Federated Learning (FL) has emerged as a promising paradigm in distributed machine learning, enabling collaborative model training while preserving data privacy. However, despite its many advantages, FL still contends with significant challenges---most notably regarding security and trust. Zero-Knowledge Proofs (ZKPs) offer a potential solution by establishing trust and enhancing system integrity throughout the FL process. Although several studies have explored ZKP-based FL (ZK-FL), a systematic framework and comprehensive analysis are still lacking. This article makes two key contributions. First, we propose a structured ZK-FL framework that categorizes and analyzes the technical roles of ZKPs across various FL stages and tasks. Second, we introduce a novel algorithm, Verifiable Client Selection FL (Veri-CS-FL), which employs ZKPs to refine the client selection process. In Veri-CS-FL, participating clients generate verifiable proofs for the performance metrics of their local models and submit these concise proofs to the server for efficient verification. The server then selects clients with high-quality local models for uploading, subsequently aggregating the contributions from these selected clients. By integrating ZKPs, Veri-CS-FL not only ensures the accuracy of performance metrics but also fortifies trust among participants while enhancing the overall efficiency and security of FL systems.

\end{abstract}


\markboth{Accepted by IEEE Communications Magazine, January 20, 2026}{Accepted by IEEE Communications Magazine, January 20, 2026}

\section{Introduction}\label{s:intro}
\IEEEPARstart{M}{achine} Learning (ML), as the cornerstone of modern AI, has evolved into a pivotal force that drives transformative innovations across various domains. As the demand for superior ML model performance grows, collaborative efforts pooling data from multiple entities have gained considerable traction. However, sharing raw data among these entities can lead to potential privacy breaches. To address this challenge, Federated Learning (FL) \cite{yang2019federated} was introduced as a privacy-preserving framework for distributed ML. Unlike traditional distributed ML, which involves sharing sample data across entities, FL transmits only model parameters, thereby safeguarding sensitive information within data. This characteristic renders FL particularly suitable for cross-institutional data collaborations in privacy-critical domains such as healthcare and finance. Furthermore, with ongoing technological advancements, FL is poised to become an essential paradigm in Internet of Things (IoT) and edge computing scenarios, facilitating more efficient and secure model training.

Despite its significant advantages, FL still faces several pressing security and trust challenges that require careful attention. For example, clients might engage in dishonest practices---such as performing fake training to free-ride on the contributions of honest participants or submitting malicious local models designed to poison the server's aggregation process.  On the other hand, the server itself might behave maliciously during aggregation, potentially disrupting FL's normal operations. For instance, a rogue server could employ inference attack strategies to extract sensitive information from the clients' submitted models \cite{pasquini2022eluding}. Another critical issue concerns the efficiency of model communication and aggregation in FL. In conventional FL frameworks, every client transmits its locally trained model to the central server, which aggregates these models into a global model during each communication round. This approach can inadvertently incorporate subpar local models, thereby degrading overall performance. Although client selection mechanisms---where the server evaluates and filters out underperforming models---offer a potential solution, they still necessitate each client transmitting its model for evaluation, leading to increased communication overhead and reduced system efficiency.


To address these challenges, zero-knowledge proof (ZKP) \cite{goldwasser2019knowledge} based federated learning (ZK-FL) has emerged as a powerful solution, particularly in cross-organizational or other high-stakes settings (e.g., finance, healthcare, and autonomous driving) where mutual trust between the server and clients is limited and strong auditability and computational integrity guarantees are required. ZK approaches specifically address threats related to client malfeasance that other privacy methods (like differential privacy or homomorphic encryption) often miss, such as free-riding attacks and model poisoning via quality verification. ZK-FL is particularly suitable for scenarios requiring verifiable correctness, hardware independence (unlike Trusted Execution Environments which rely on specific hardware), and communication efficiency.

The contribution of this paper is twofold, as summarized below:
\begin{itemize}[label=$\bullet$]
	\item While ZKP can be integrated into FL to bolster trust among participants, research on ZK-FL is still in its early stages, and there is a lack of a systematic classification and comprehensive summary of the related work. In this paper, we thus first introduce a structured ZK-FL framework that categorizes the technical roles of ZKP within each stages of FL (e.g., local training, model evaluation, aggregation), and we classify, analyze, and summarize existing ZK-FL studies accordingly.
	
	\item Based on the structured ZK-FL framework, we then propose a novel FL algorithm that applies ZKP during the client selection process within the ``model evaluation'' stage, named the Verifiable Client Selection FL (Veri-CS-FL) algorithm. In this approach, while local models are trained across various participating clients, cosine similarity metrics between these models and a benchmark model are used for evaluation. The benchmark model is derived by training on the central server using its proprietary data. During metric generation, ZKP is employed to produce proofs that verify the accuracy of the metric calculation. The server then only selects and aggregates the local models with higher-ranking, validated metrics. Lower-ranking, poorly performing local models no longer need to be uploaded to the server. By incorporating ZKP, our approach not only enhances the evaluation and communication efficiencies of local models through transparent evidence of their quality but also fosters increased trust among clients.
\end{itemize}


The rest of this article is organized as follows. Section \ref{s:bg} provides the background knowledge of FL and ZKP. Section \ref{s:3} summarizes ZK-FL. Section \ref{s:4} gives the model system and the overall framework for the proposed algorithm. Section \ref{s:PERFORMANCE EVALUATION} discusses the system tests and the conclusion is presented in Section \ref{s:conclusion}.

\section{Background}\label{s:bg}

This section presents the technical background related to ZK-FL, summarizing the foundational principles of FL and ZKP.

\subsection{Federated Learning}\label{ss:2a}

In a FL system, there are $K$ participating clients, with each client $C_k$ maintaining a local training dataset $D_k$, $k\in{1,2,\cdots,K}$, alongside a central server $S$ that functions as an aggregator to facilitate the distributed and collaborative training of a ML model using the local data from these clients. The entire model training process in FL is divided into multiple iterative rounds, conducted sequentially.

In the $t$-th round of FL, the server $S$ first distributes the latest parameters of the global model $w^{t-1}$ to all clients. Then, each client $C_k$ trains a local model $w_k^t$ based on the global model $w^{t-1}$ using their own local dataset $D_k$. After that, all clients transmit their trained local models $\{w_k^t,k=1,2,\cdots,K\}$ to the server $S$. Finally, the server $S$ obtains a new global model $w^{t}$ updated in the $t$-th round by aggregating the clients' local models according to some predefined aggregation rule. This process repeats until the global model converges or a predetermined number of iterations is reached.

Existing FL algorithms often operate under the assumption that both clients and the server are inherently trustworthy, an assumption that may not hold in practical distributed scenarios. Addressing the challenges of safeguarding data privacy and mitigating the impact of Byzantine nodes during the FL process is therefore critical. In this context, \textit{Byzantine nodes} refer to participants that behave arbitrarily or maliciously---ranging from sending incorrect model updates to intentionally disrupting the training protocol---posing a severe threat to system reliability. This article focuses on the zero-knowledge proof based strategies proposed to prevent malicious behaviors from both servers and clients.

\subsection{Zero-Knowledge Proof}\label{ss:2b}
ZKP is a cryptographic technique that proves a statement's validity without revealing any private information about the statement. Among several types of ZKP algorithms, zero-knowledge succinct non-interactive argument of knowledge (zk-SNARK) is considered to be the most practical. A zk-SNARK algorithm is usually represented by an arithmetic circuit that consists of the basic arithmetic operations of addition, subtraction, multiplication, and division. The three algorithmic components of a zk-SNARK algorithm are given:

\begin{itemize}[label=$\bullet$]
    \item $(\mathsf{PK}, \mathsf{VK}) \leftarrow \mathsf{KEYGEN}(1^\lambda, R)$ is the key generation algorithm that generates the proving key $\mathsf{PK}$ and the verification key $\mathsf{VK}$ by using a predefined security parameter $\lambda$ and an arithmetic circuit representation $R$.
    \item $\pi \leftarrow \mathsf{PROVE}(\mathsf{PK}, x, W)$ is the proof generation algorithm that generates a proof $\pi$ based on the proving key $\mathsf{PK}$, the input $x$ and the witness $W$.
    \item $b \leftarrow \mathsf{VERIFY}(\mathsf{VK}, x, \pi)$ is the proof verification algorithm that outputs a binary decision to accept ($b=1$) or reject ($b=0$) $\pi$ using $\mathsf{VK}$, $x$ and $\pi$ as the input.
\end{itemize}

The proving key $\mathsf{PK}$ and the verification key $\mathsf{VK}$ generated by the $\mathsf{KEYGEN}$ algorithm are treated as the public parameters pre-generated by an authority. The $\mathsf{PROVE}$ algorithm is executed by the prover, and the $\mathsf{VERIFY}$ algorithm is executed by the verifier. Witness $W$ is the secret owned by the prover that he/she does not want to reveal to others and yet wants to prove that he/she knows the secret. The technical advantages of zk-SNARK include its short proof size, fast proof verification, and non-interactive proving. In this article, we consider using zk-SNARK to generate zero-knowledge proofs to provide trust for our Veri-CS-FL algorithm. From a privacy perspective, the zero-knowledge property of zk-SNARK ensures that the generated proof $\pi$ reveals nothing about the witness $W$ other than the validity of the statement being proven. This cryptographic guarantee prevents any information leakage regarding the prover's private data during the verification process.

\section{ZK-FL Framework} \label{s:3}

\begin{figure*}[!t]
    \centering
    \includegraphics[width=16cm]{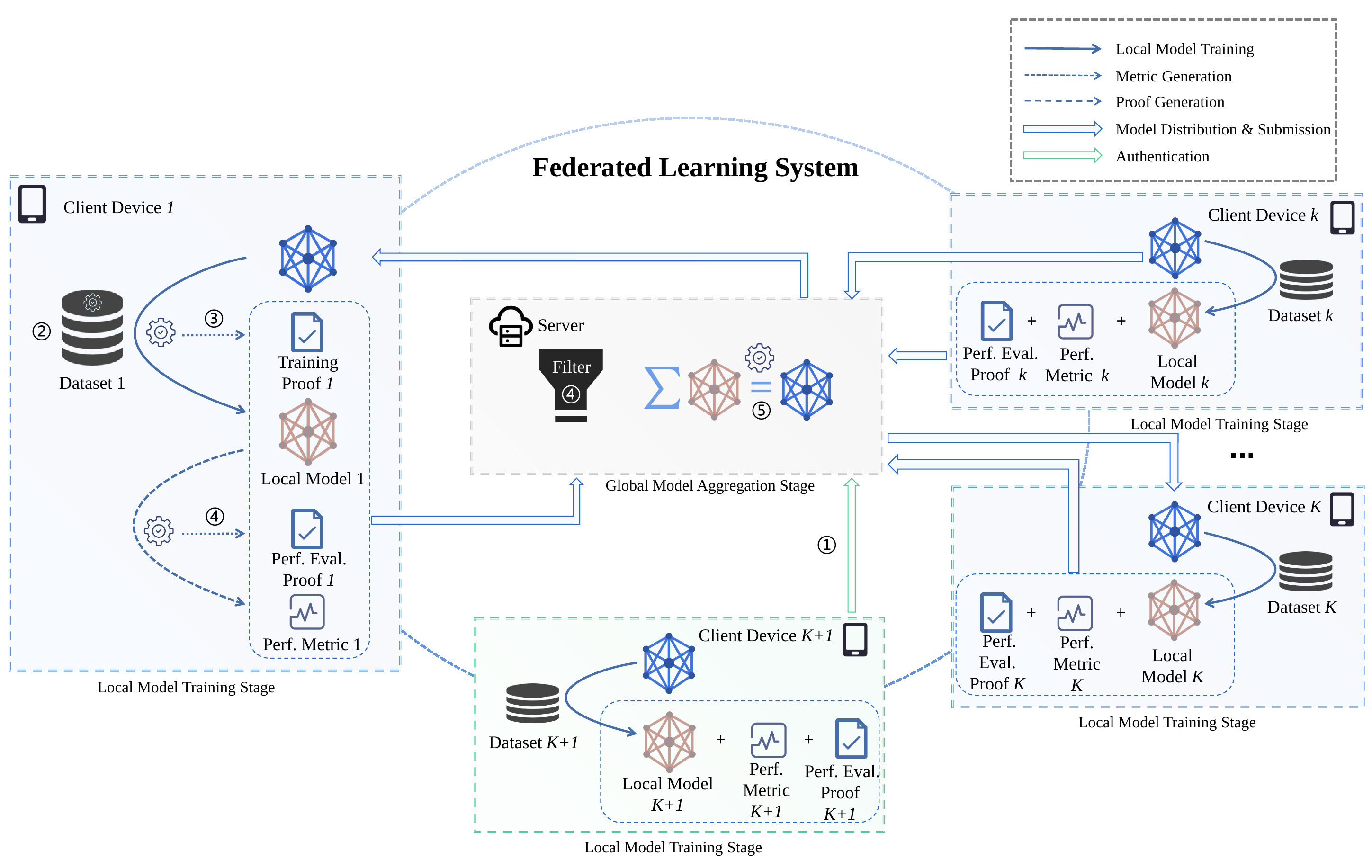}
    \caption{The structured framework of ZK-FL demonstrating that ZKP can be integrated at various stages of FL to address different trust issues in FL: Local Model Training Stage: \textcircled{1} client identity trust, \textcircled{2} local data trust, and \textcircled{3} local training trust; Global Model Aggregation Stage: \textcircled{4} model evaluation trust, and \textcircled{5} model aggregation trust.}
    \label{f2}
\end{figure*}

In this section, we analyze the potential trust issues that may arise at various stages of the FL process and categorize them into five distinct types. For each trust issue, we provide corresponding ZKP solutions, combining the analysis of trust issues and ZKP solutions into a unified structured framework. Our structured framework reflects the current state of ZK-FL research and provides guidance for future advancements. The structured framework of ZK-FL is depicted in Fig. \ref{f2} and will be further elaborated upon in the following discussions.

Trust in FL refers to the expectation that participating entities (clients and servers) will act as intended regarding data, models, and computational processes. Trust issues arise when this expectation is violated, leading to attacks that compromise the system's security. These issues occur at two key stages: local model training and global model aggregation, as demonstrated in Fig. \ref{f2}. Below, we analyze these trust issues and discuss how ZKP can solve them.

\subsection{Local Model Training Stage}
During local model training, trust must be established to ensure the authenticity of participating clients, the quality of their local data, and the integrity of their training processes. ZKP can be employed to address the following trust issues:

\subsubsection{Client Identity Trust}

\begin{itemize}
    \item {\bf Trust Issue}: Untrusted clients may gain access to the FL system and disrupt the training process by submitting harmful or fake models.

    \item {\bf ZKP Solution}: ZKP can verify client identities without exposing sensitive information. For instance, \cite{ji2023lafed} introduces a lightweight authentication framework for blockchain-based FL, using ZKP to verify client identities.
\end{itemize}

\subsubsection{Local Data Trust}

\begin{itemize}
    \item {\bf Trust Issue}: Clients may use fake or low-quality data for training, deceiving the server and other clients.

    \item {\bf ZKP Solution}: ZKP can ensure the authenticity and integrity of local training data by requiring clients to create cryptographic commitments to their data and generate proofs that the committed data was used for training the model \cite{abbaszadeh2024zero}. This approach preserves data confidentiality while ensuring reliability.
\end{itemize}

\begin{table*}[!t]
	\centering
	\caption{Summary of the Existing ZK-FL Solutions.}
	\label{tab:my_label}
	\begin{tabularx}{\textwidth}{c p{3.5cm} X p{3.5cm}}
		\toprule
		\textbf{Stage} & \textbf{Trust Issue} & \textbf{Solution} & \textbf{Reference} \\
		\midrule
		
		\multirow{5}{*}{\makecell{Local Model \\ Training Stage}} 
		& Client Identity Trust & \textcircled{1} Ensuring the Authenticity of Client Identities & \cite{ji2023lafed} \\ 
		\cmidrule(l){2-4}
		& Local Data Trust & \textcircled{2} Ensuring the Integrity and Accuracy of the Training Data & \cite{abbaszadeh2024zero} \\ 
		\cmidrule(l){2-4}
		& Local Training Trust & \textcircled{3} Ensuring the Integrity and Validity of Model Updates & \cite{heiss2022advancing}, \cite{wang2023enhancing}, \cite{ma2024vpfl},\cite{duan2024verifiable} \\ 
		\midrule
		
		\multirow{2}{*}{\makecell{Global Model \\ Aggregation Stage}} 
		& Model Evaluation Trust & \textcircled{4} Ensuring Integrity of Performance Metric Evaluations & \cite{ruckel2022fairness} and this work \\ 
		\cmidrule(l){2-4}
		& Model Aggregation Trust & \textcircled{5} Ensuring Trustworthy Model Aggregation & \cite{ma2024trusted}, \cite{wang2024zkfl}, \cite{11204667} \\ 
		\bottomrule
	\end{tabularx}
\end{table*}

\subsubsection{Local Training Trust}

\begin{itemize}
    \item {\bf Trust Issue}: Even with authenticated clients and high-quality data, some clients may bypass predefined training protocols, submit low-quality models, or even introduce malicious models to disrupt the system.

    \item {\bf ZKP Solution}: ZKP can verify that local training processes adhere to predefined protocols. Works such as \cite{heiss2022advancing, wang2023enhancing, ma2024vpfl, duan2024verifiable} leverage ZKPs to verify protocol-compliant local computation, e.g., proving that reported updates are computed from the prescribed model and training procedure (rather than being fabricated or arbitrarily manipulated), while preserving the confidentiality of local data.
\end{itemize}

\subsection{Global Model Aggregation Stage}
At the global model aggregation stage, trust issues arise in the evaluated performances of local models and the aggregation process itself. ZKP can address these challenges as follows:

\subsubsection{Model Evaluation Trust}

\begin{itemize}
    \item {\bf Trust Issue}: The server must use the evaluated performances of local models to filter out submitted low-quality local models. However, self-reported performance evaluation results from clients may be inaccurate or misleading. A malicious server also may intentionally provide biased or inaccurate evaluations of local models to serve its own agenda, undermining the fairness and reliability of the FL process.

    \item {\bf ZKP Solution}: ZKP can prove the correctness of performance metrics evaluated on the client side. For example, \cite{ruckel2022fairness} employs ZKP to ensure reliable performance evaluations. Our proposed approach (discussed in the next section) also leverages ZKP to prove and verify client-reported performance metrics, enabling the selection of high-performing models for aggregation while filtering out poorly performing ones.
\end{itemize}

\subsubsection{Model Aggregation Trust}

\begin{itemize}
    \item {\bf Trust Issue}: The server may deviate from predefined aggregation protocols, introduce biases, or unfairly weigh client contributions. From the clients' perspective, there is often a lack of transparency in the aggregation process.

    \item {\bf ZKP Solution}: ZKP can ensure trustworthy aggregation by providing verifiable evidence that the aggregation procedure follows the agreed rule and that malformed contributions are detectable. Works such as \cite{ma2024trusted, wang2024zkfl} demonstrate how ZKPs can be integrated into FL to enable verifiable aggregation under Byzantine behaviors (including decentralized settings), improving transparency and integrity of the aggregation outcome. Additionally, \cite{11204667} proposes a Byzantine-robust FL framework that embeds ZKPs to ensure verifiable aggregation even under a malicious server, addressing threats in dynamic and non-IID scenarios.
\end{itemize}

Based on the locations where trust issues occur at different stages of the FL process, we have marked them with the corresponding numbers in Fig. \ref{f2}. By integrating ZKP into FL, these key trust issues can be solved, ensuring the reliability of various FL operations. ZKP techniques facilitate verifiable and transparent processes, thereby enhancing the overall trustworthiness of FL systems. Table \ref{tab:my_label} summarizes the existing ZK-FL solutions, highlighting their respective contributions according to this proposed ZK-FL framework.

\section{Our ZK-FL Solution}\label{s:4}

The previous section provides a systematic review of potential threats in FL and corresponding ZKP-based solutions. This section focuses on trust issues in model performance evaluation, describing the system threats and presenting our ZK-FL solution to that. 

\subsection{Threat and Trust Model}

We consider a threat and trust model for FL with the following setup: 

\begin{itemize}
    \item There is no inherent trust between the server and the participating clients.
    \item The server, which is responsible for model aggregation, is assumed to be trustworthy and reliable, while the clients may exhibit malicious behavior.
    \item Malicious clients can engage in poisoning attacks by submitting unreliable models with poor performance, aiming to disrupt the overall learning process.
\end{itemize}

Under this threat and trust model, updating the global model with poorly performing local models uploaded from malicious clients could severely compromise FL system performance. To mitigate this risk, the server must access each uploaded model's performance and select those local models that meet predefined performance standards to aggregate. This process is referred to as {\bf client selection}. While a straightforward method for client selection is to have clients report their model performance metrics and then select, this self-reporting is unreliable due to the inherent mistrust between the server and clients, as well as the risk of malicious behavior. Consequently, the server should independently evaluate each client's local model before selection. However, this approach introduces two significant drawbacks:

\begin{itemize}
    \item[1.] It places an additional computational burden on the server. When the model being trained is large and the number of clients is substantial, this additional computation overhead when evaluating multiple local model at the server may become unacceptable.

    \item[2.] It leads to inefficient use of communication bandwidth between the server and the clients. If certain local models are found to exhibit poor performance and are subsequently excluded from aggregation, the communication bandwidth consumed during the transmission of these models is effectively wasted. This issue becomes even more pronounced when the model size is large and communication takes place over wireless channels, resulting in high costs.
\end{itemize}

In the following, we propose a novel ZK-FL solution that leverages ZKP to implement model evaluation and client selection in FL (Veri-CS-FL) that does not have the above drawbacks. 


\subsection{Veri-CS-FL}

In this section, we propose our ZK-FL algorithm, i.e., the Veri-CS-FL algorithm, to achieve the trusted model evaluation at the client-side and client selection at the server-side without re-evaluating the local models submitted by the clients for verifications. Veri-CS-FL is also based on the FedAvg algorithm framework, and it leverages local models' performance metrics as the criterion to select the clients' local models with satisfied performances for aggregation. Rather than re-running all the clients' local models to compute their performance metrics at the server side, Veri-CS-FL integrates ZKP to validate the accuracy of the computed performance metrics in a communication-and-computation-efficient and privacy-preserving way. The specific processes of Veri-CS-FL are described as follows:

On the server side, Veri-CS-FL designates a small and clean dataset $D_s$ as the root dataset, enabling the server $S$ to train a benchmark model to evaluate the performances of the clients' local models. At the start of the FL training process, the key generation algorithm of zk-SNARK is executed at a trusted third party to generate the proving key ($\mathsf{PK}$) for all clients, who act as provers, and the verification key ($\mathsf{VK}$) for the server, who acts as the verifier: $(\mathsf{PK}, \mathsf{VK}) \leftarrow \mathsf{KEYGEN}(1^\lambda, R)$, where $R$ denotes the circuit that models the computation process for evaluating model performance. At the start of the $t$-th FL training round, all entities, including the server and the clients, access the global model, $w^{t-1}$, trained in the previous FL training round. Then, the following training steps are conducted between the server and the clients in the $t$-th FL training round:

\begin{itemize}
    \item[1.] The server $S$ utilizes the root dataset $D_s$ to train a benchmark model $w_s^t$ from the latest global model $w^{t-1}$. The server $S$ broadcasts $w_s^t$ to all the participating clients to provide a model benchmark.

    \item[2.] Each client $C_k$ trains an updated local model $w_k^t$ from the last round global model $w^{t-1}$ utilizing its local dataset $D_k$, $k\in{1,2,\cdots,K}$.\footnote{We assume that the local model training stage is conducted honestly by all clients, which is further guaranteed through a ZKP algorithm, as demonstrated in \cite{heiss2022advancing, wang2023enhancing, ma2024vpfl, duan2024verifiable}. We only consider the model evaluation trust in the design of VeriCS-FL.} Subsequently, each client computes a performance metric of the updated local model $w_k^t$ by comparing the updated local model $w_k^t$ with the benchmark model $w_s^t$, and employs the proof generation algorithm of zk-SNRAK to generate a zero-knowledge proof $\pi_k$ for the performance metric of $w_k^t$: $\pi_k \leftarrow \mathsf{PROVE}(\mathsf{PK},x,W)$, where the public input $x$ is the benchmark model of the server $w_s^t$ together with the hash of the updated local model $h_k^t=\text{hash}(w_k^t)$: $x = \left\{ {w_s^t,h_k^t} \right\}$, and the witness $W$ is the updated local model $w_k^t$: $W=w_k^t$.

    \item[3.] Each client $C_k$ transmits the proof $\pi_k$ for the updated local model's performance metric and the hash of their updated local model $h_k^t$ to the server $S$. Instead of transmitting the full model immediately, clients first transmit the proof. Only selected clients upload the model subsequently.

    \item[4.] The server $S$ collects the performance metrics of the clients' local models and their corresponding proofs, $\pi_k$, $k\in{1,2,\cdots,K}$. The server first verifies all the proofs using the proof verification algorithm of zk-SNARK: $ \mathsf{VERIFY}(\mathsf{VK}, x, \pi)$ with $x = \left\{ {w_s^t,h_k^t} \right\}$ and  $\pi=\pi_k$, and proceeds to rank the corresponding performance metrics whose proofs have been successfully verified. The performance metrics whose proofs cannot pass the verification are not used and ranked. The server $S$ then selects the top $N$ clients based on their ranked performance metrics into the candidate set $\mathcal{C}$ and notifies the respective clients in this set to upload their updated local models.

    \item[5.] The selected $N$ clients in the candidate set $ \mathcal{C}$  transmit their updated local models, $w_k^t$, $C_k \in \mathcal{C}$, to the server $S$.

    \item[6.] The server $S$ collects the updated local models sent from the $N$ clients in the candidate set $\mathcal{C}$. The server first computes the hash of the updated local model sent by each selected client in step 5: $\widetilde h_k^t=\text{hash} (w_k^t)$, and compares the computed hash with the hash, $h_k^t$, sent by the client in step 3. If the two hashes match, the server $S$ will labels this client as a valid candidate; if the two hashes differ, the server $S$ will remove this client $C_k$ from the candidate set $ \mathcal{C}$. After this hash verification operation for each client $C_k \in \mathcal{C}$, the server $S$ aggregates these updated local models of the clients still in the candidate set $\mathcal{C}$ according to the predetermined rule to derive the updated global model $w^t$ for the $t$-th round and then distribute $w^t$ to all the clients.
\end{itemize}

The above processes of Veri-CS-FL iterate continuously until the final global model converges or the predetermined number of epochs is reached. Some remarks about Veri-CS-FL are given below.
\begin{algorithm}[!t]
\caption{Verified Client Selection Federated Learning (Veri-CS-FL)}
\label{algorithm}
\begin{algorithmic}[1]
\REQUIRE
Each client $C_{k}$ possesses the proving key $\mathsf{PK}$ of zk-SNARK and the local dataset $D_{k}$ , $\forall k \in \{1,2,\cdots,K\}$ while the server $S$ holds the verifying key $\mathsf{VK}$ of zk-SNARK and the small and clean dataset $D_s$. 
\ENSURE
Global model $w^{T}$.
\STATE Server $S$ sends $w^0$ to initialize the global model on client $C_k$, $\forall k \in \{1,2,\cdots,K\}$.

\FOR{iteration $t=1,2,...,T$}

\STATE \textbf{Server $S$:}

\STATE \hspace{0.7em} uses $D_s$ to train the benchmark $w_s^t$ from $w^{t-1}$;
\STATE \hspace{0.7em} broadcasts $w_s^{t}$ to all the participating clients.
\vspace{0.1cm}
\FORALL{\textbf{Client} $C_k$, $\forall k \in \{1,2,\cdots,K\}$ in parallel}
\STATE trains the local model $w_k^t$ from $w^{t-1}$ based on $D_{k}$;
\STATE computes \text{cos}$_k^t$ by comparing $w_k^t$ and $w_s^{t}$;
\STATE generates $\pi_k^t$ of $w_k^t$: $\pi_k^t \leftarrow \mathsf{PROVE}(\mathsf{PK},x,W)$ with $h_k^t=\text{hash}(w_k^t)$, $x = \left\{ {w_s^t,h_k^t} \right\}$ and  $W=w_k^t$;
\STATE transmits $\pi_k^t$ and $h_k^t$ to the server $S$.
\ENDFOR
\vspace{0.1cm}
\STATE \textbf{Server $S$:}
\STATE \hspace{0.7em} sorts \text{cos}$_k^t$, $\forall k \in \{1,2,\cdots,K\}$;
\STATE \hspace{0.7em} verifies all the proofs: $b \leftarrow \mathsf{VERIFY}(\mathsf{VK}, x, \pi)$ with $x = \left\{ {w_s^t,h_k^t} \right\}$ and  $\pi=\pi_k$;
\STATE \hspace{0.7em} if the proof $\pi_k^t$ is verified ($b=1$), the corresponding \text{cos}$_k^t$ will be ranked, and the top $N
$ will be selected;
\STATE \hspace{0.7em} if \text{cos}$_k^t$ is selected, the corresponding client will be put in the candidate set $\mathcal{C}$ and notified.
\vspace{0.1cm}
\FORALL{\textbf{Client} $C_k$, $\forall k \in \{1,2,\cdots,N\}$, in $\mathcal{C}$}
\STATE transmits $w_k^t$ to the server $S$.
\ENDFOR
\vspace{0.1cm}
\STATE \textbf{Server $S$:}
\STATE \hspace{0.7em} computes $\widetilde h_k^t=\text{hash} (w_k^t)$;
\STATE \hspace{0.7em} compares $\widetilde h_k^t$ with $h_k^t$ (line 9): if $\widetilde h_k^t \neq h_k^t$, removes the corresponding $C_k$ from $\mathcal{C}$, $\forall k \in \{1,2,\cdots,N\}$.
\STATE \hspace{0.7em} aggregates all the $w_k^t$ of clients in $\mathcal{C}$ to obtain $w^t$.
\STATE \hspace{0.7em} derives $w^t$ to all the clients.
\ENDFOR
\STATE return $w^{T}$
\end{algorithmic}
\end{algorithm}

\begin{itemize}
    \item[1.] In Veri-CS-FL, each client needs to submit the performance metric of their local model to the server. The server then uses the performance metrics to select local models with satisfied performances, and uses the training sample sizes as the weights in the weighted aggregation method of global models. There are many different performance metrics that can be used to characterize the performance of a local model. For example, the loss function of the model on the test dataset, the KL divergence of the output predicted by the model on the test dataset. In Veri-CS-FL, we propose to adopt the cosine similarity between the benchmark model and the local model \cite{liang2023auditable} as the performance metric:   $ \text{cos}_k^t\triangleq{\text{cos}}\left(w_k^t,w_s^t \right) = \frac{w_k^t \cdot w_s^t}{|w_k^t||w_s^t|}$. When utilizing the benchmark model as a reference, the larger the cosine similarity between the local model and the benchmark model, the better the performance of the local model. The computation of cosine similarity can also be easily translated to the arithmetic circuit of zk-SNARK, enabling the computation of the zero-knowledge proof for it.
    
    \item[2.] By generating the zero-knowledge proofs for the performance metrics of local models at the client-sides, and validating the proofs at the server side, Veri-CS-FL can achieve the following advantages: i) Communication-efficient: the local models with poor performances will not be communicated between the clients and the server. This can save much uplink communication bandwidth, particularly when the models are large. ii) Computation-efficient: the server has no need to re-run all the local models to evaluate their performances. The verification of zero-knowledge proofs in zk-SNARK is very efficient, e.g., the zk-SNARK algorithm of Groth16 \cite{groth2016size} has a fixed verification complexity and fixed proof size regardless of the sizes of public inputs and private witness. iii) Trusted collaboration: The client transmits the performance metrics and its zero-knowledge proofs to the server, and if the server feels that its updated local model has satisfactory performance, it asks the client for the updated local model. As a result, the client and server can agree on the model training contribution. It prevents cases where after the client passes the updated local model to the server, the server refuses to admit that the client provided a local model with satisfactory performance in this round of training, and maliciously devalues the actual contribution of the client.
 
\end{itemize}

Algorithm~\ref{algorithm} provides the pseudo-codes of Veri-CS-FL. In the next section, we will experimentally evaluate Veri-CS-FL.

\subsection{Security and Privacy Analysis}
In this subsection, we analyze the security properties of Veri-CS-FL, focusing on completeness, soundness, zero-knowledge, and resistance to poisoning attacks.

\subsubsection{Completeness}
For any honest client $C_k$ that correctly executes the local training and metric calculation protocol, the generated proof $\pi_k$ will always pass the verification $\mathsf{VERIFY}(\mathsf{VK}, x, \pi_k) = 1$. This ensures that legitimate contributions are always recognized by the server.

\subsubsection{Soundness}
The soundness property ensures that a malicious client cannot generate a valid proof for a false statement. Specifically, if a client submits a high performance metric but has not actually trained a model that yields such a metric (or has not trained a model at all), it is computationally infeasible to generate a proof $\pi$ that satisfies the verification algorithm. This prevents clients from free-riding or falsifying their contributions.

\subsubsection{Zero-Knowledge}
The zero-knowledge property guarantees that the proof $\pi_k$ reveals no information about the witness $W$ (the local model weights $w_k^t$) other than the validity of the statement (the cosine similarity metric). The server learns only the metric and the hash of the model, ensuring that the model parameters remain private until the client is selected for aggregation. Notably, the only information exposed is a single scalar value (the cosine similarity score), which is insufficient to reconstruct the high-dimensional model parameters or infer the underlying training data distribution. The zero-knowledge property preserves the privacy of FL. 

\subsubsection{Resistance to Poisoning Attacks}
Veri-CS-FL inherently resists model poisoning attacks through its verifiable client selection mechanism. By filtering models based on their cosine similarity to a benchmark model trained on clean data, the system effectively excludes malicious updates that deviate significantly from the benign optimization direction. Unlike standard FL, where the server aggregates all models (including potentially poisoned ones), Veri-CS-FL verifies the quality \textit{before} aggregation.

\subsubsection{Impact on Heterogeneous Data}
A potential concern is whether filtering based on a benchmark model restricts the learning of unique features from outlier datasets. Veri-CS-FL mitigates this by using cosine similarity, which measures the alignment of the update direction rather than the specific parameter values. Valid updates from non-IID clients, even if they contain unique features, generally align with the global optimization direction (positive similarity). In contrast, malicious or purely noisy updates typically exhibit orthogonality or opposition. Furthermore, by selecting the top $N$ contributors rather than applying a rigid threshold, the system adapts to the available pool of updates, preserving the most constructive contributions even in heterogeneous environments.

\section{Experimental Investigation}
\label{s:PERFORMANCE EVALUATION}

In this section, we conduct numerical experiments on the Veri-CS-FL algorithm that we proposed in Section IV to investigate the use of ZKP for the model evaluation and client selection in FL.

\subsection{Simulation Settings}


We conduct numerical experiments based on two image datasets: MNIST and FEMNIST. The MNIST dataset is used to evaluate Veri-CS-FL under independent and identically distributed (IID) settings, while the FEMNIST dataset, which is inherently partitioned by writer, is employed to evaluate the algorithm under non-IID settings, reflecting more realistic heterogeneous data distributions across clients. The FL process involves 20 clients, in which each client has 3000 sample data injected with Gaussian noise at 4 different levels. Meanwhile, the server $S$ possesses a clean and small dataset of 400 sample data as its root dataset. During training, the used loss function is the symmetric cross entropy, which combines cross-entropy and inverse cross-entropy to address the over-fitting problem. As discussed in last section, the top $N$ local models are selected for aggregation after their performances' ZKP is verified. Therefore, for the aggregation process, we set $N$ as 4, 8, 12, 16, and 20 respectively.

To investigate the computation complexity of  ZKP generation and verification, and the size of proof, we set the model size (the length of the vectors $w^t_k$) to values from the set $\{8, 16, 32, 64, 128, 256, 512, 1024, 2048, 4096\}$. The zk-SNARK algorithm of Groth16 \cite{groth2016size} is used to generate and verify ZKP in Veri-CS-FL. We implement the Groth16 program for Veri-CS-FL using the library of ZoKrates (https://zokrates.github.io/). All experiments were conducted on a server equipped with an Intel Xeon CPU and 64GB RAM. The reported proof generation and verification times include the core proving time of the Groth16 backend, where ZoKrates compiles the high-level code into R1CS (Rank-1 Constraint System) constraints for measurement.

\subsection{Results Analysis}

\begin{figure*}[!th]
    \centering
    \captionsetup{justification=centering}
    \subfigure{
    \includegraphics[width=0.48\textwidth]{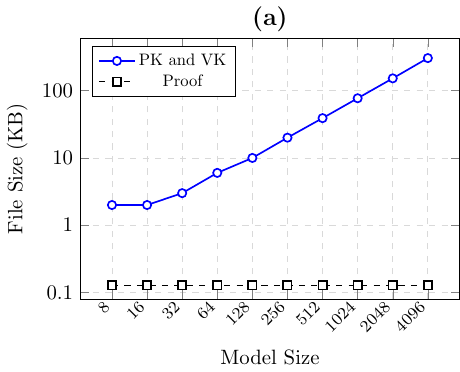}
    \label{fig:result1}
    }
    \subfigure{
    \includegraphics[width=0.45\textwidth]{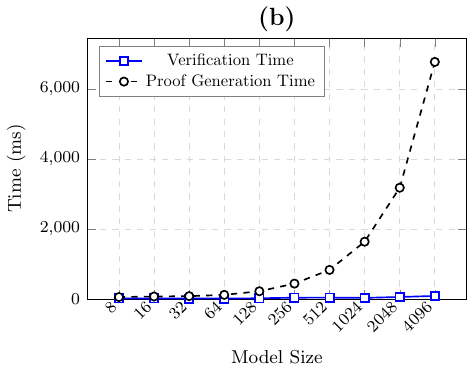}
    \label{fig:result2}
    }
    \subfigure{
    \includegraphics[width=0.48\textwidth]{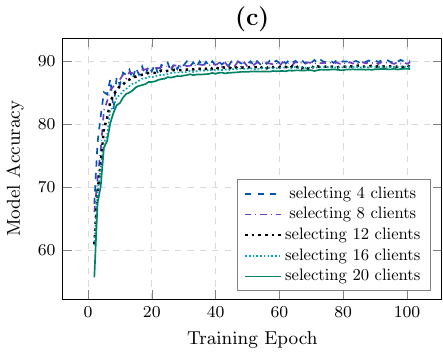}
    \label{fig:result3}
    }
    \subfigure{
    \includegraphics[width=0.42\textwidth]{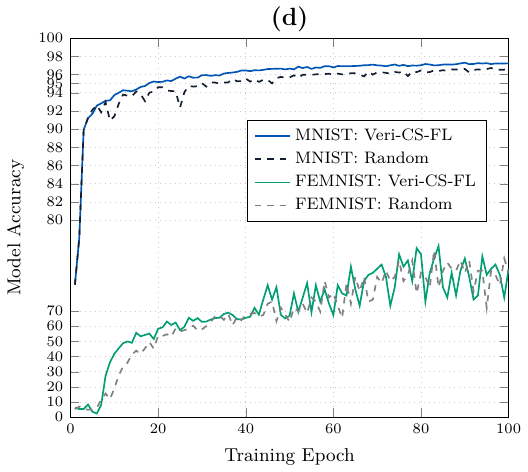}
    \label{fig:result4}
    }
    \caption{Experimental results of Veri-CS-FL: (a) Proof size with respect to model size; (b) Verification time and proof generation time with respect to model size; (c) Global model accuracy when aggregating local models from different numbers of clients; (d) Global model accuracy when using performance-based client selection (Veri-CS-FL) and random client selection (Rand-CS-FL). 
    }
 
    \label{fig:enter-label}
  
\end{figure*}

We investigate the impact of model size on the proof size, proof generation time and proof verification time. As illustrated in Fig. \ref{fig:result1}, the sizes of the proving and verification keys increase linearly with the model size. This is because these keys encapsulate the circuit constraints required for the proof problem, and the size of these constraints grows proportionally with the model size. Notably, the proof size remains constant at 128 bytes, a characteristic feature of Groth16. In Fig. \ref{fig:result2}, it is evident that the verification time remains constant across varying model sizes, owing to the inherent verification efficiency of Groth16. In contrast, the proof generation time increases significantly with larger model sizes. However, for moderate model sizes, this increase remains within a practical and acceptable range.

We then investigate the performance of client selection in FL by training a model with $10^3$ parameters. In Fig. \ref{fig:result3}, we illustrate the accuracies of the global models aggregated in Veri-CS-FL under varying numbers of selected clients. The results demonstrate that as the number of selected clients decreases from 20 to 4, the global model's performance consistently improves, peaking at over 90\%. This improvement is likely due to the noise present in individual clients' models---aggregating a larger number of noisy models can degrade overall performance. To benchmark the client selection performance of Veri-CS-FL, we also simulate an alternative FL scheme, Rand-CS-FL, where the server randomly selects $N$ clients from all the client to submit their local models, rather than selecting clients based on their models' performance. We selected Rand-CS-FL as the baseline instead of aggregating all models to ensure a fair comparison under identical communication constraints (selecting top $N$ clients), reflecting realistic edge scenarios where bandwidth is limited. In Fig. \ref{fig:result4}, we compare the performances of Veri-CS-FL and Rand-CS-FL with $N=4$. The results show that Veri-CS-FL outperforms Rand-CS-FL by effectively identifying clients with high-performing local models for aggregation, thereby improving the global model's overall performance.

To validate the robustness of Veri-CS-FL on non-IID data scenarios, we conducted additional experiments using the FEMNIST dataset, which is explicitly designed for FL in non-IID settings. The number of selected clients is $N=4$. The root dataset includes 400 sample data randomly sampled from the whole dataset. The results, reported in  Fig. \ref{fig:result4}, confirm that our client selection strategy remains effective in such scenarios, successfully filtering out low-quality updates and improving convergence speed compared to random selection. Specifically, in the non-IID FEMNIST case, Veri-CS-FL significantly accelerates early convergence by using the benchmark to filter out conflicting gradients caused by data heterogeneity. In contrast, for the IID MNIST case, Veri-CS-FL primarily excels in the later fine-tuning stages by filtering out noisy updates to reach a higher accuracy ceiling.

Regarding scalability, Veri-CS-FL scales efficiently. Server-side verification of Groth16 proofs is constant-time ($\approx$ milliseconds) and embarrassingly parallel, allowing the server to handle thousands of clients. Client-side proof generation is performed locally, ensuring horizontal scalability as the number of clients increases.

In terms of communication overhead, a Groth16 proof is approximately 128 bytes, and a SHA-256 hash is 32 bytes, totaling a negligible $\approx$ 160 bytes per client. In contrast, preventing the upload of even a single poorly performing model (which can be megabytes or gigabytes in size) yields bandwidth savings that far outweigh the cost of transmitting proofs. Thus, the overhead is $O(1)$ while the savings are $O(\text{Model Size})$. This reduction in communication overhead enhances the effectiveness of Veri-CS-FL in bandwidth-constrained scenarios (e.g., wireless edge networks) or high-security contexts, where efficient data transmission and minimized communication costs are essential. Moreover, rapid advancements in hardware acceleration for ZKPs (e.g., on mobile GPUs/NPUs) are making client-side proof generation increasingly feasible, further enhancing the practicality of Veri-CS-FL on resource-constrained devices.

\section{Conclusion}\label{s:conclusion}

In this article, we construct a systematic ZK-FL framework that classifies and examines the technical roles of ZKP across various FL modules. Through this framework, we provide a structured analysis of current ZK-FL approaches, identifying key areas where ZKP enhances both trust and efficiency in FL systems. Additionally, we propose the Verifiable Client Selection FL (Veri-CS-FL) algorithm, a novel approach that leverages ZKP to verify local model metrics based on their similarity to a benchmark model. By integrating ZKP, Veri-CS-FL ensures the reliable selection of high-quality models, improves model aggregation, and fosters trust among participants. While this work primarily addresses the upstream communication bottleneck, future research may explore integrating downstream compression techniques or selective model broadcasting to further reduce the total communication overhead in the FL lifecycle.

\bibliographystyle{IEEEtran}
\bibliography{ref.bib}

\end{document}

%% file: packageSetup.tex
\usepackage{graphicx}
\usepackage{cite}
\usepackage{picinpar}
\usepackage{amsmath}
\usepackage{url}
\usepackage{flushend}
\usepackage{colortbl}
\usepackage{soul}
\usepackage{multirow}
\usepackage{pifont}
\usepackage{color}
\usepackage{alltt}
\usepackage{hyperref}
\usepackage{enumerate}
\usepackage{epstopdf}
\usepackage{pbox}
\usepackage{enumitem}
\usepackage{fancyhdr}

\usepackage{bm}
\usepackage{amssymb}
\usepackage{stfloats}
\usepackage{algorithmic}
\usepackage[ruled,vlined]{algorithm2e}
\usepackage[table]{xcolor}